%% file: main.tex
\documentclass[conference]{IEEEtran}
\IEEEoverridecommandlockouts
\input{preamble}

\begin{document}

\ifthenelse{\boolean{isPreprint}}{%
    \title{Accurate GPU Memory Prediction for Deep Learning Jobs through Dynamic Analysis} 
}{
    \title{xMem: A CPU-Based Approach for Accurate GPU Memory Prediction in Deep Learning Training}
}


\author{
    \ifthenelse{\boolean{isPreprint}}{%
        \IEEEauthorblockN{Jiabo Shi}
        \and
        \IEEEauthorblockN{Yehia Elkhatib}
    }{
        \IEEEauthorblockN{Anonymous}
    }
    
}

\maketitle

\begin{abstract}

\ifthenelse{\boolean{isPreprint}}{%
    The benefits of Deep Learning (DL) impose significant pressure on GPU resources, particularly within GPU cluster, where Out-Of-Memory (OOM) errors present a primary impediment to model training and efficient resource utilization. 
    Conventional OOM estimation techniques, relying either on static graph analysis or direct GPU memory profiling, suffer from inherent limitations: static analysis often fails to capture model dynamics, whereas GPU-based profiling intensifies contention for scarce GPU resources. 
    To overcome these constraints, \projectname emerges. It is an innovative, entirely CPU-based analysis tool capable of accurately predicting the peak GPU memory required for DL training tasks without accessing the target GPU. This "offline" prediction capability is \projectname's core advantage, allowing accurate memory footprint information to be obtained before task scheduling, thereby effectively preventing OOM and optimizing GPU allocation. 
    Its performance was validated through thousands of experimental runs across convolutional neural network (CNN) models: compared to baseline GPU memory estimators, \projectname significantly reduces the relative error by 84\%\@\xspace, lowers the estimation failure probability by 73\%\@\xspace. \projectname represents a key step towards efficient and predictable DL training in resource-constrained environments.

}{
    The widespread adoption of Deep Learning (DL) in diverse application areas has significantly increased the demand for GPUs. Consequently, GPU resources are scarce and are managed in clusters to maximize resource utilization. 
    However, this shift introduces new debugging challenges when training DL models on shared clusters 
    particularly Out-Of-Memory (OOM) errors, an issue commonly reported in industry and academic literature. 
    Existing solutions for avoiding OOM primarily rely on static analysis of the DL model's computational graph, or leverage GPU resources directly or indirectly to estimate the peak memory required for training the given task on the target GPU. 
    Unfortunately, relying on GPUs for these predictions exacerbates resource contention and increases scheduling challenges. Furthermore, the dynamic nature of model development limits the accuracy of static analysis to 
    estimate peak memory usage. 
    To address these limitations, we propose \projectname, a novel tool that uses CPU-based analysis to accurately predict the memory required for model training on a GPU. By eliminating the reliance on GPUs for memory estimation, \projectname promotes efficient GPU utilization while mitigating OOM errors. 
    Our empirical evaluation of 16 DL models (a total of \xmemtotalevaruns runs) demonstrates that, compared to state-of-the-art GPU memory estimators, \projectname decreases the median relative error by \xmemmedianerrorimprove, reduces the average probability of estimation failure by \xmemprobabilityimprove, accelerates the runtime by \xmemruntimeimprove, and improves memory conservation by \xmemmemoryimprove.
}

\end{abstract}

\begin{IEEEkeywords}

\ifthenelse{\boolean{isPreprint}}{%
    Deep Learning,
    Memory Management, 
    Program Analysis, 
}{
    Deep Learning,
    PyTorch, 
    Memory Management, 
    Memory Optimization, 
    Program Analysis, 
    GPU Efficiency
}
\end{IEEEkeywords}

\section{Introduction}
\label{introduction/chapter}

\subfile{introduction}

\section{Background}
\label{background/chapter}

\subfile{background}


\section{Design}
\label{design/chapter}

\subfile{design}

\section{Evaluation}
\label{evaluation/chapter}

\subfile{evaluation}


\section{Related Work}
\label{related-work/chapter}
\subfile{related_work}

\section{Conclusion}
\label{conclusion/chapter}
\subfile{conclusion}

\bibliographystyle{IEEEtran}
\bibliography{refs-edit}

\end{document}

%% file: preamble.tex
\usepackage[moderate,tracking=normal]{savetrees} 

\usepackage[pdfcreator={}, pdfproducer={}]{hyperref}
\usepackage{url}
\usepackage{hyperref}
\usepackage[ruled,vlined]{algorithm2e}
\usepackage{subfiles}
\usepackage{cite}
\usepackage{amsmath,amssymb,amsfonts}
\usepackage{ifthen}
\usepackage{graphicx}
\usepackage{subcaption}
\usepackage{textcomp}
\usepackage[table]{xcolor}
\definecolor{stripe}{RGB}{254,254,217}
\definecolor{exp-xMem}{HTML}{377eb8}
\definecolor{exp-DNNMem}{HTML}{e41a1c}
\definecolor{exp-SchedTune}{HTML}{4daf4a}
\definecolor{exp-LLMem}{HTML}{984ea3}
\definecolor{fq-bl}{HTML}{2C7BB6}
\definecolor{fq-br}{HTML}{FDAE61}
\definecolor{fq-tl}{HTML}{ABD9E9}
\definecolor{fq-tr}{HTML}{D7191C}

\def\BibTeX{{\rm B\kern-.05em{\sc i\kern-.025em b}\kern-.08em
    T\kern-.1667em\lower.7ex\hbox{E}\kern-.125emX}}
\usepackage{colortbl}
\usepackage{array}
\usepackage{booktabs}
\newcommand{\midsepremove}{\aboverulesep=0mm \belowrulesep=0mm}
\midsepremove
\newcommand{\midsepdefault}{\aboverulesep=0mm \belowrulesep=0mm}
\midsepdefault

\usepackage[inline]{enumitem}
	\setlist[enumerate]{leftmargin=1.2em}
	\setlist[itemize]{leftmargin=1.2em,label={$\bullet$}}%

\newboolean{isPreprint}
\setboolean{isPreprint}{true}

\ifthenelse{\boolean{isPreprint}}{%
    \newcommand{\projectname}{VeritasEst\@\xspace}
    \newcommand{\imagePath}{}
}{
    \newcommand{\projectname}{xMem\@\xspace}
    \newcommand{\imagePath}{assets/}
}
\newcommand{\xmemtotalevaruns}{5,040\@\xspace}
\newcommand{\xmemprobability}{13.59\%\@\xspace}
\newcommand{\xmemmemory}{4.55 GB\@\xspace}
\newcommand{\xmemmedianerror}{5.46\%\@\xspace}
\newcommand{\xmemruntime}{11.72s\@\xspace}

\newcommand{\xmemprobabilityimprove}{73.44\%\@\xspace}
\newcommand{\xmemmedianerrorimprove}{84.32\%\@\xspace}
\newcommand{\xmemmemoryimprove}{125.36\%\@\xspace}
\newcommand{\xmemruntimeimprove}{50.16\%\@\xspace}

\usepackage{xspace}

\newcommand*{\cf}{\textit{cf.}\@\xspace}
\newcommand*{\eg}{\textit{e.g.,}\@\xspace}
\newcommand*{\ie}{\textit{i.e.,}\@\xspace}
\makeatletter
\newcommand*{\etc}{%
    \@ifnextchar{.}%
        {etc}%
        {etc.\@\xspace}%
}
\newcommand*{\etal}{%
    \@ifnextchar{.}%
        {et al}%
        {et al.\@\xspace}%
}
\makeatother

%% file: introduction.tex
The field of Deep Learning (DL) is witnessing rapid growth, 
with Deep Neural Network (DNN) models finding broad application across diverse sectors, including autonomous driving \cite{grigorescu_survey_2020}, speech recognition \cite{xiong_microsoft_2018}, and image generation \cite{wang_cnn-generated_2019}, where they have demonstrated performance levels comparable to those of humans in various domains \cite{guerriero_iterative_2023}. Consequently, technology companies are increasing investments in high-performance computing infrastructure, particularly Graphics Processing Units (GPUs) to accelerate model training. This surge in demand has created notable GPU scarcity; even leading AI companies, such as OpenAI, are facing challenges in securing adequate GPU resources \cite{raza_openais_2023}. 

Furthermore, companies use computing clusters to optimize the utilization of GPUs. However, contrary to expectations, debugging tasks that should ideally be handled locally are also being transferred to clusters due to the absence of local GPU support. 
Avoiding potential runtime out-of-memory (OOM) issues is a key focus of these debugging efforts.
Previous studies \cite{cheng_towards_2023, zhang_empirical_2020} provide evidence that GPU clusters in Microsoft and Meta are experiencing OOM problems, with 9\% of DL training tasks failing due to OOM. This problem further exacerbates the scarcity of GPUs. Using expensive GPUs for OOM error detection is highly inefficient and wasteful. 

Recently, strategies for dealing with OOM problems in clusters fall primarily into two categories. The first strategy, which includes approaches like AntMan \cite{258957}, employs offloading techniques \cite{li_cotrain_2023, ren_zero-offload_2021} to move tensors that exceed the maximum memory capacity of the GPU to the host RAM. 
However, this technique has a limitation: permanent partial memory on RAM affects training performance when the task's memory consumption exceeds the device's maximum capacity, and it does not fully address the OOM problem.
The second strategy, which we are focusing on, aims to address the OOM problem by predicting the peak GPU memory usage in advance, \eg \cite{gao_estimating_2020, albahar_schedtune_2022, liu_tbem_2022, yeung_horus_2022, kim_llmem_2024}, to determine whether the requested GPU can meet the training requirements, essentially resolving the OOM issue before it occurs, and helping the clusters conserve GPU resources.

\ifthenelse{\not\boolean{isPreprint}}{%
    \begin{figure}[tbp]
        \centering
        \includegraphics[width=1\linewidth]{\imagePath/Motivation-Memory Change Impacted by zero out operation.pdf}
        \caption{Segment and Tensor memory usage changes are greatly influenced by where the zeroing gradient function, \texttt{optimizer.zero\_grad}, is placed. Position 0 (POS0) means calling it just before backward propagation, while Position 1 (POS1) means calling it at the beginning of each iteration.}
        \label{motivation/fig/zero-out-memory-change}
    \end{figure}
}

Solutions for estimating peak GPU memory face three main challenges. 
The first key challenge is understanding when memory is being allocated and freed during a training task. Previous works, like Gao et al. \cite{gao_estimating_2020} and Liu et al. \cite{liu_tbem_2022}, focused on static analysis of the computational graph but faced limitations due to the lack of dynamic and detailed memory tracking, especially for operations like gradient zeroing and altering the optimizer. 
\ifthenelse{\not\boolean{isPreprint}}{%
    Fig.~\ref{motivation/fig/zero-out-memory-change} shows that even minor adjustments in the placement of the zeroing gradient function from position 0 to position 1 within a training loop can considerably affect memory usage patterns.
}

Second, a frequent misunderstanding in DL model training is the assumption that the tensor memory footprint directly reflects GPU memory usage. In PyTorch, the default memory allocator for CUDA, CUDACachingAllocator \cite{pytorch_cuda_2024}, is used for memory management. This allocator can cause differences between the sizes of the tensors and the actual memory allocated to the GPU. Consequently, to accurately assess peak GPU memory usage, it is necessary to consider the allocator's behavior rather than considering only the tensor sizes.

The third challenge is whether GPU memory estimation can be performed without relying on expensive GPU hardware, which makes it feasible to perform estimation on any server without access to specialized hardware.

In this paper, we propose \textit{\projectname}, a GPU memory estimator that can utilize CPU-based profiling data to infer the peak memory required for training a model on a GPU. This estimator is based on three principal observations: 
\begin{enumerate*}[label=(\roman*)]
    \item the execution sequence of high-level code, such as Python, remains consistent across CPU and GPU, differing merely in the low-level kernel implementation, such as ATen \cite{pytorch_cpp_2024};
    \item the optimized operators in recent versions of PyTorch demonstrate comparable memory usage on both CPU and GPU; and
    \item consistent memory allocation and deallocation sequences yield predictable outcomes across CPU and GPU when managed by the same allocator.
\end{enumerate*}

\ifthenelse{\boolean{isPreprint}}{%
    We implemented \projectname and evaluated its performance extensively in a total of thousands of experiment runs. We evaluated \projectname with convolutional neural network (CNN) models, five optimizers, varying batch sizes, resulting in a \xmemmedianerror relative error, and a \xmemprobability probability of estimation failure. 
    Compared with three state-of-the-art GPU memory estimators \cite{gao_estimating_2020, albahar_schedtune_2022, kim_llmem_2024}, \projectname decreases the relative error by \xmemmedianerrorimprove, the probability of estimation failure by \xmemprobabilityimprove.

}{
    We implemented \projectname and evaluated its performance extensively in a total of \xmemtotalevaruns experiment runs. We evaluated \projectname with 16 models, five optimizers with batch sizes ranging from 10 to 1,000, and two gradient zeroing positions, resulting in a \xmemmedianerror median relative error, a \xmemprobability probability of estimation failure, an average run time \xmemruntime, and an average \xmemmemory of memory conserved from each execution. 
    Compared with three state-of-the-art GPU memory estimators \cite{gao_estimating_2020, albahar_schedtune_2022, kim_llmem_2024}, \projectname decreases the median relative error by \xmemmedianerrorimprove, the probability of estimation failure by \xmemprobabilityimprove, the runtime by \xmemruntimeimprove, and improves memory conservation by \xmemmemoryimprove.
}

This study makes the following contributions.
\ifthenelse{\boolean{isPreprint}}{%
    \begin{enumerate}
        \item We propose a novel solution that can not only accurately dynamically estimate GPU peak memory consumption for DL model training tasks, but also provide additional memory change trace during training, requiring only a CPU (\S\ref{design/chapter}). Additionally, \projectname's is open source. 
        \item We offer a GPU memory sequence of model training tasks for all memory simulators by performing a thorough analysis of CPU-based profiling data generated by the PyTorch Profiler (\S\ref{design/chapter/seq-orchestration}).
        
    \end{enumerate}
    \noindent
}{
    \begin{enumerate}
        \item We propose a novel solution that can not only accurately dynamically estimate GPU peak memory consumption for DL model training tasks, but also provide additional memory change trace during training, requiring only a CPU (\S\ref{design/chapter}). Additionally, \projectname's is open source. 
        \item Our approach has been verified to demonstrate significant capability for GPU memory conservation by preventing OOM issues and providing a minimum runnable memory for each training job (\S\ref{evaluation/chapter/monte-carlo/memory-saving}).
        \item We offer a GPU memory sequence of model training tasks for all memory simulators by performing a thorough analysis of CPU-based profiling data generated by the PyTorch Profiler (\S\ref{design/chapter/seq-orchestration}).
        \item We developed a Python simulator for the CUDACachingAllocator, which helps users understand the memory traces of the CUDACachingAllocator without requiring a GPU (\S\ref{design/chapter/memort-allocator}).
        
    \end{enumerate}
    \noindent
    We provide the \projectname source code, baselines, and full results {\href{https://anonymous.4open.science/r/ICDCS-2025-xMem-9017}{online}.\label{introduction/foot/source-code}}

}

%% file: background.tex
\subsection{Out-of-Memory Issues} \label{background/chapter/OOM-issue}


Previous studies \cite{cheng_towards_2023, zhang_empirical_2020} have shown that GPU clusters at both Microsoft and Meta are experiencing OOM issues, which cause approximately 9\% of deep learning training tasks to fail due to OOM. Additionally, an independent empirical study that examined 2,716 Stack Overflow posts identified OOM as one of six major deep learning error-related issues \cite{islam_comprehensive_2019}. OOM challenges are common in the field of system resource management, particularly concerning DL training, due to the limited GPU memory available compared to the high memory demands of model training. Several factors, including batch size, choice of the optimizer, and even small code alterations, can affect memory usage during training. The studies \cite{gao_estimating_2020, liu_tbem_2022, yeung_horus_2022} proposed static analysis methods to estimate the maximum GPU memory usage by utilizing a computation graph or static data extracted from a model. However, developing and training models is not static but rather dynamic, with the aim of optimizing the model's performance by tuning the code. A dynamic GPU memory estimator is a handy tool for preventing OOM or efficient resource management.

\ifthenelse{\not\boolean{isPreprint}}{%
    \begin{figure}[tbp]
        \centering
        \includegraphics[width=1\linewidth]{\imagePathbk-memory consumption.png}
        \caption{Memory Management Diagram, showing how memory blocks are managed in PyTorch for a training job.}
        \label{background/fig/memory-management}
    \end{figure}
}


\ifthenelse{\not\boolean{isPreprint}}{%
    \begin{algorithm}[tbp]
        \small
    \caption{Calculate segment size}
    \label{background/alg/segment-size-calculation}
    \KwIn{Requested size $size$}
    \KwOut{Allocation size}
    $kSmallSize \gets 1MB$\;
    $kSmallBuffer \gets 2MB$\;
    $kMinLargeAlloc \gets 10MB$\;
    $kLargeBuffer \gets 20MB$\;
    $kRoundLarge \gets 2MB$\;
    
    \If{$size \leq kSmallSize$}{
        \Return $kSmallBuffer$\;
    }
    \ElseIf{$size \leq kMinLargeAlloc$}{
        \Return $kLargeBuffer$\;
    }
    \Else{
        \Return $kRoundLarge \times \left( \frac{size + kRoundLarge - 1}{kRoundLarge} \right)$\;
    }
    \end{algorithm}
}

\subsection{Memory Consumption}
Typically, there are two main types of memory consumption in a PyTorch training job: the memory consumed by Tensors and the memory consumed by a memory allocator, termed Segments.

\subsubsection{Tensors} \label{background/chapter/tensor-memory}
refer to memory blocks used to store tensors during model training or inference in PyTorch, including model parameters, activations, gradients, etc. During forward and backward propagations, the tensors consume memory and their memory requirements are influenced by factors such as model architecture, batch size, data precision, etc. Notably, the tensor memory blocks are not directly allocated from a GPU. Instead, tensor memory blocks are allocated from segments that are blocks of GPU memory previously allocated and cached by the allocator.  
\ifthenelse{\not\boolean{isPreprint}}{%
    As depicted in Fig.~\ref{background/fig/memory-management}, the solid green rectangle denotes the tensor memory blocks. Furthermore, if the tensor does not inherently satisfy this specification, its size is rounded to the nearest power of 512.
}

\subsubsection{Segments} \label{background/chapter/segment}
are memory blocks that are requested from GPU and managed by the CUDACachingAllocator. Segments are relatively larger blocks, and all memory requests from DL model training tasks actually request a fitted memory block from it. The allocator retains the deallocated memory blocks used by tensors within segments for reuse instead of releasing it back to the GPU. When segment memory is insufficient, a new segment is requested from the GPU. 
\ifthenelse{\not\boolean{isPreprint}}{%
    Fig.~\ref{background/fig/memory-management} illustrates the segments as rectangles with a yellow crisscrossed pattern, where the size of these segments is determined by an underlying algorithm implemented within the allocator, as depicted in Algorithm~\ref{background/alg/segment-size-calculation}. 
}
Thus, GPU memory consumption is denoted as the total memory size of segments requested by the allocator. All previous studies \cite{yeung_horus_2022, albahar_schedtune_2022, kim_llmem_2024} overlook the importance of a memory allocator to estimate GPU memory consumption during training.

\subsection{Sequence of Memory Activities}
The sequence of memory allocation and deallocation activities is a critical factor in determining memory consumption. Different sequences may induce varying levels of fragmentation, thereby directly influencing the overall memory usage of deep learning models. 
\ifthenelse{\not\boolean{isPreprint}}{%
    The order of these memory activities significantly affects the ability to predict GPU memory requirements with precision, as illustrated in Fig.~\ref{background/fig/memorg-impact-by-sequence}. It exhibits two different memory sequences with segment sizes determined using Algorithm~\ref{background/alg/segment-size-calculation}. The only difference between Sequence 1 and Sequence 2 is that in Sequence 2, the deallocation activity follows the allocation of 2 MB, whereas in Sequence 1, the deallocation follows immediately after its corresponding allocation activity, while all other operations remain the same. The results indicate that Sequence 1 consumed 196 MB of GPU memory, whereas Sequence 2 consumed only 118 MB. This slight modification in sequence led to a substantial difference in memory consumption, with Sequence 1 using approximately 67.80\% more memory than Sequence 2. 
}
As mentioned in Section~\ref{background/chapter/OOM-issue}, the studies \cite{gao_estimating_2020, liu_tbem_2022} consistently utilize static memory sequences, whereas the research studies \cite{yeung_horus_2022, albahar_schedtune_2022, kim_llmem_2024} did not take into account of memory sequences in their memory estimation processes.



\ifthenelse{\not\boolean{isPreprint}}{%
    \begin{figure}[tbp]
        \centering
        \includegraphics[width=1\linewidth]{\imagePath/bk-memory sequence.png}
        \caption{The impact on GPU memory consumption by memory sequence activities}
        \label{background/fig/memorg-impact-by-sequence}
    \end{figure}
}

%% file: design.tex
\ifthenelse{\boolean{isPreprint}}{%
    \begin{figure}[btp]
    \centering
        \begin{minipage}{.45\textwidth}
            \centering
            \includegraphics[width=1\linewidth]{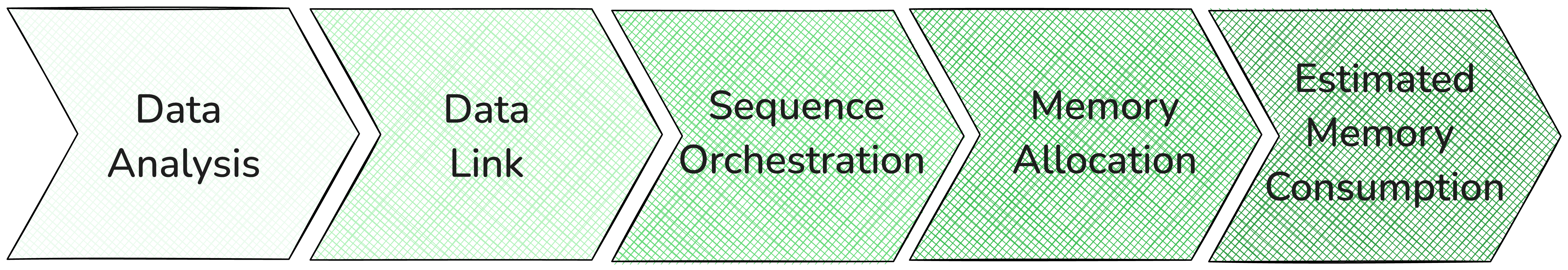}
            \caption{Workflow of \projectname Memory Estimator.}
            \label{design/fig/workflow}
            \vspace{1em}
            \centering
            \includegraphics[width=1\linewidth]{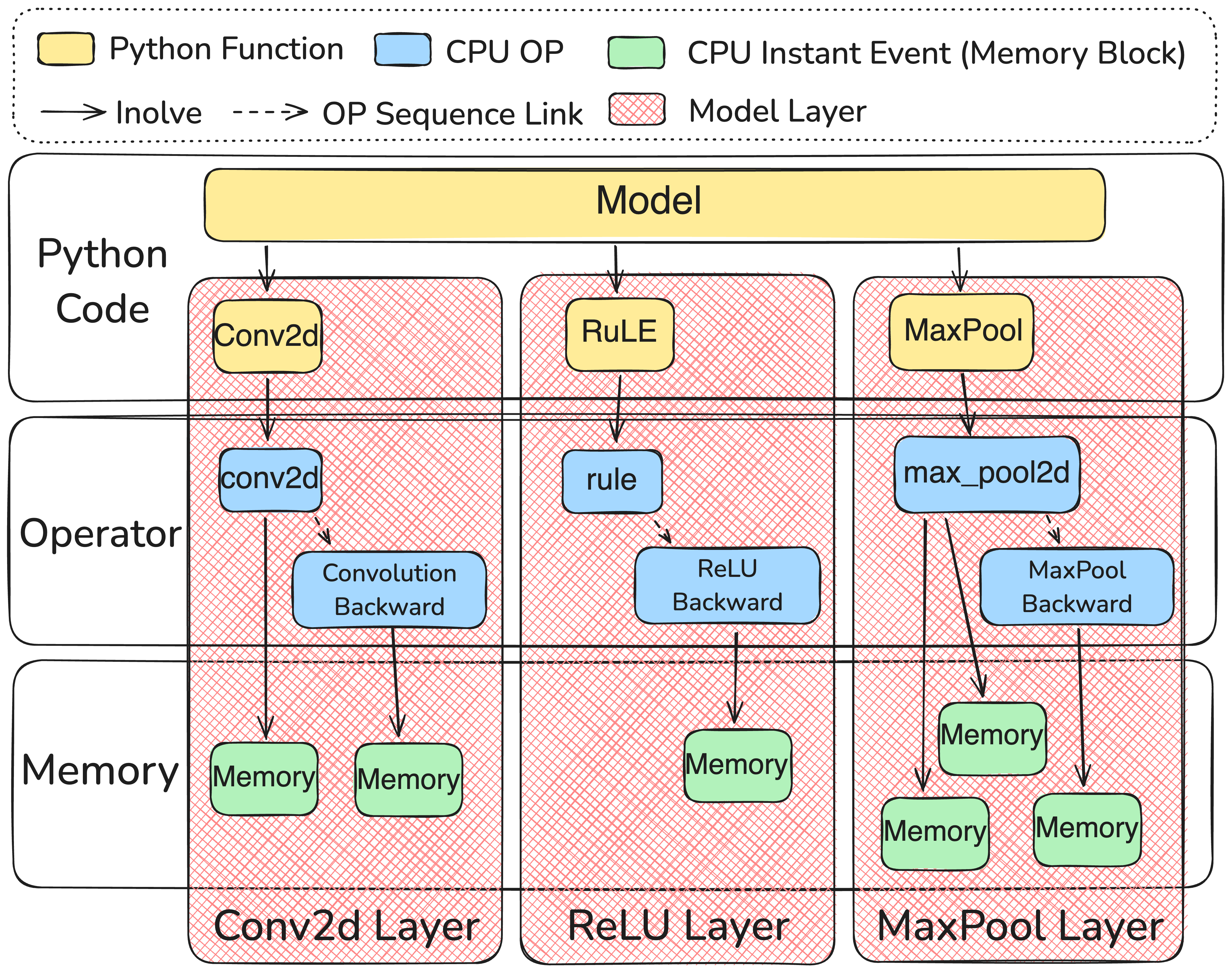}
            \caption{A call hierarchical structure diagram showing links between memory blocks, operators, and layers.}
            \label{design/fig/data-linking}
        \end{minipage}%
    \end{figure}
}{
    \begin{figure*}[tbp]
    \centering
        \begin{minipage}{.45\textwidth}
            \centering
            \includegraphics[width=1\linewidth]{\imagePath/Design-xMem Workflow.png}
            \caption{Workflow of \projectname Memory Estimator.}
            \label{design/fig/workflow}
            \vspace{1em}
            \centering
            \includegraphics[width=1\linewidth]{\imagePath/Design-mechanism of memory orchestraction.png}
            \caption{The mechanism of memory block arrangement in the sequence orchestration section.}
            \label{design/fig/memory-orchestraction}
        \end{minipage}%
        \hspace{2em}
        \begin{minipage}{0.45\textwidth}
            \centering
            \includegraphics[width=1\linewidth]{\imagePath/Design-data analysing step-linking.png}
            \caption{A call hierarchical structure diagram showing links between memory blocks, operators, and layers.}
            \label{design/fig/data-linking}
        \end{minipage}
    \end{figure*}

}

As GPU characteristics are expensive and scarce, we proposed \projectname, a CPU-based memory estimator that uses a CPU that is anywhere ready to profile training tasks, enabling prediction of GPU memory requirements and eliminating the need for GPUs during neither development nor execution. Its major objective is to accurately predict peak GPU memory usage, referred to as the minimum runnable GPU memory, during DL model training on GPUs, with the goal of preventing OOM issues and optimizing GPU memory usage. Moreover, the current version is specific for PyTorch \cite{paszke_pytorch_2019}, which is a well-known DL framework that is widely used in both research and industry.

Due to their distinct hardware architectures, memory allocation functions differ significantly when a DL model is trained on CPUs versus GPUs. To address this challenge, we have developed techniques to extrapolate from CPU profiling to accurately estimate GPU requirements based on a set of key principles outlined below:
\begin{enumerate}
    \item \textbf{Sequence}: PyTorch must ensure compatibility with various processors also known as \textit{kernels}, such as CPU, GPU, and XPU. To achieve this, it employs a dispatch mechanism that separates the implementation in high-level languages like Python from the actual execution code \cite{pytorch_cpp_2024} within the kernel, thus maintaining their independence. Therefore, the function call sequence during an identical DL training task remains consistent between CPU and GPU.
    \item \textbf{Memory}: All fundamental C++ operators in PyTorch are optimized to manage memory efficiently regardless of the target kernel.\footnote{Such as $\text{aten::mkldnn\_convolution}$ for CPU kernels and \text{aten::cudnn\_convolution} for CUDA kernels.} 
    This means that the memory usage for most operators are expected to be comparable, if not identical, for training on either the CPU or GPU.  
    \item \textbf{Allocation}: A memory allocator handles all real memory allocation activities. Whenever any operator requires to acquire or free a memory block, the associated function in the allocator is invoked. This indicates that memory allocation can be fully predictable when a properly ordered memory input is provided.
\end{enumerate}

Building on the aforementioned principles, \projectname follows the workflow shown in Fig.~\ref{design/fig/workflow}. 
Initially, the data analysis phase (\S\ref{design/chapter/data-analysis}) ensures that all data are prepared for subsequent use. 
Next, the data link phase (\S\ref{design/chapter/data-link}) identifies all memory activities necessary for each layer of the model. 
The sequence orchestration phase (\S\ref{design/chapter/seq-orchestration}) then corrects the timing of the memory activities to align them more closely with the GPU memory activities. 
\ifthenelse{\boolean{isPreprint}}{%
    In the final step, the memory allocation phase ensures that all memory blocks are adjusted to the correct rounded-up sizes and allocated to the appropriate segments. Moreover, it also guarantees that all segments are computed to the designed size by an algorithm before requesting the size from the GPU. 
}{
    In the final step, the memory allocation phase (\S\ref{design/chapter/memort-allocator}) ensures that all memory blocks are adjusted to the correct rounded-up sizes and allocated to the appropriate segments. Moreover, it also guarantees that all segments are computed to the designed size by an algorithm before requesting the size from the GPU. 
}

\ifthenelse{\boolean{isPreprint}}{%
    The allocator is highly adaptable and can be replaced by any allocator defined by the user. In this study, we use the Python implementation of the CUDACachingAllocator simulator with the Best Fit with Coalescing (BFC) algorithm \cite{hasan_study_2005} as the default memory allocation method.
}{
    The allocator is highly adaptable and can be replaced by any allocator defined by the user. In this study, we use the Python implementation of the CUDACachingAllocator simulator with the Best Fit with Coalescing (BFC) algorithm \cite{hasan_study_2005} as the default memory allocation method, and Algorithm~\ref{background/alg/segment-size-calculation} for determining the segment size.
}

\subsection{Data Analysis} \label{design/chapter/data-analysis}

\begin{algorithm}[tbp]
    \small
\caption{\texttt{cpu\_instant\_event} Grouping}
\label{design/alg/memory-activities-grouping}
\KwData{cpu\_instant\_event Data}
\KwResult{Set of Time-based Memory Block}

$addr\_map \gets \{\}$\;
$node\_map \gets \{\}$\;
$data \gets \text{get\_sorted\_cpu\_event\_data()}$\;

\ForEach{$trace \in data$}{
    $addr \gets \text{get\_addr}(trace)$\;

    \eIf{$addr \notin addr\_map$}{
        $block \gets \text{create\_memory\_block}(trace)$\;
        $addr\_map[addr] \gets block$\;
    }{
        $addr\_map[addr].\text{mark\_as\_free}(trace)$\;
        $block \gets addr\_map.\text{pop}(addr)$\;
        $node\_map[block.\text{alloc\_time}].\text{append}(block)$\;
    }
}
\ForEach{$remaining \in addr\_map$}{
    $alloc\_time \gets remaining.\text{alloc\_time}$\;
    $node\_map[alloc\_time].\text{append}(remaining)$\;
}
\Return{$\text{sort\_by\_alloc\_time}(node\_map)$}\;
\end{algorithm}

Our primary data source is the log of events created by the PyTorch Profiler \cite{pytorch_pytorch_2024}. This phase ensures that all data are structured in specific formats that are conducive to sequential analysis. From the 24 types of events provided in JSON format, we rely on four event types only, as follows.

\subsubsection{\texttt{python\_function}}
\label{design/chapter/python-function}
provides linkage details, such as parent and current Python function IDs, helping to construct a function call graph to identify a hierarchy within the training task. In particular, the traces of these Python functions clearly indicate the PyTorch layers that are invoked, such as VGG16, Conv2d, or ReLU. As \texttt{python\_function} events contain clear linking details, they can readily be arranged into a tree structure, preserving only the information related to the PyTorch layer in the data structure.

\subsubsection{\texttt{cpu\_op}}
\label{design/chapter/cpu-op}
provides extensive metadata regarding operators, encompassing input size, input type, and sequence number. The sequence number is the pivotal attribute in this additional information because it is used to link operators that are created during forward and backward propagation of the same layer. Unlike \texttt{python\_function}, \texttt{cpu\_op} does not furnish direct linkage metadata that would assist in constructing a stack tree. Therefore, we employ the fundamental concept of the interval tree algorithm to organize time-based CPU operators, with only the root nodes of the operators used after it. 

\subsubsection{\texttt{user\_annotation}}
\label{design/chapter/user-annonation}
is designed to spotlight specific code in profiling data. In PyTorch, the official source code marks key codes such as \texttt{profiler.step}, \texttt{optimizer.zero\_grad}, and \texttt{optimizer.step}. The \texttt{profiler.step} marks the start of a new profiling cycle, facilitating analysis by breaking down profile data into multiple iterations. The \texttt{optimizer.step} initiates a parameter update, marking the beginning of updates after all propagation. \texttt{optimizer.zero\_grad} helps to locate a time point when the gradients from the current or previous iteration are zeroed.

\subsubsection{\texttt{cpu\_instant\_event}}
\label{design/chapter/cpu-instant-event}
presents memory allocation and deallocation metadata, including memory address, size, total allocated, and total reserved. We refer to this as ``memory activity". It is crucial to record the lifecycle of each memory block carefully from profiling data in order to provide accurate memory sequence data to the allocator. The key task is to establish correct associations between the allocation and the deallocation memory activities. Each single memory address could be reused multiple times throughout the training process; therefore, an association based on addresses alone is not feasible. Thus, we sequentially analyze all memory activities and apply Algorithm~\ref{design/alg/memory-activities-grouping} to bind each memory activity accurately. This process generates a set of new memory blocks, which details allocation time, deallocation time, size in bytes, etc. Each new block connects two memory activities, one for allocation and one for deallocation. A block remains permanent if it has only one activity.

\subsection{Data Link} \label{design/chapter/data-link}

The linking phase determines all required memory blocks for each layer within the model by constructing a call hierarchy. As shown in Fig.~\ref{design/fig/data-linking}, this procedure begins with Python functions (yellow blocks) accessing the memory blocks (green blocks) through intermediary operators (blue blocks). Notably, memory blocks are associated only when they are generated at the execution time of each operator of each model layer. Due to the lack of a straightforward link between any two types of metadata, we rely exclusively on timestamps to correlate Python functions, operators, and memory blocks.

To establish a connection between \texttt{python\_function} and \texttt{cpu\_op}, the start and end times of each layer, excluding wrapper layers\footnote{Such as a bottleneck layer in ResNet \cite{he_deep_2016}.} are utilized to determine all operators generated in that time period. Some operators in the search results may have a sequence number attribute, indicating that they are associated with one or more operators involved in gradient computation during the backward propagation phase. Because the connections between the forward and backward operators have already been established during data analysis, 
we can easily include the corresponding backward operators within the layer and store them as part of the backward operators. If multiple sequence numbers appear in the search output, all operators linked to those sequence numbers are included.

The link between \texttt{cpu\_op} and the memory blocks generated during the data analysis phase is established in the same way as described in the preceding paragraph. 
Furthermore, by analyzing the memory operations of each model's layer on both the CPU and GPU, we determine that the memory retained is identical for each layer in the CPU and GPU. However, the temporary memory allocations within the CPU training task for the same layer exceed those in the GPU. 
Due to optimization issues in layer function implementations on the CPU, the layer function may create an additional variable to store data passed by its caller and delete the original one. This creates an extra memory block during this value-assigning process, causing memory fluctuations. To address this, 
we determine if a memory block's deallocating time falls within the execution timeframe of an operator. If so, we can filter those blocks that are allocated and freed within the operator execution time range. We can, thus, focus on the memory blocks that remain after an operator finishes.

\subsection{Sequence Orchestration} \label{design/chapter/seq-orchestration}
The objective of the orchestration phase is to construct a memory sequence that is closer to that on the GPU by aligning and rearranging memory allocation and deallocation timings. 
\ifthenelse{\not\boolean{isPreprint}}{%
    Fig.~\ref{design/fig/memory-orchestraction} shows how memory orchestration is performed in \projectname, and the four adjustments are described below.
}

\subsubsection{Model} \label{design/chapter/sequence/model}
refers to memory blocks that are loaded during a model transfer stage, which is triggered by \texttt{model.to(kernel)}. However, this part of memory information is not contained in CPU-based profiling data. Therefore, we assume that during the model transfer phase, the overall memory consumption is equivalent to the memory used by the gradients. This assumption has been validated through experiments that analyze memory data obtained from the PyTorch Snapshot Profiler \cite{pytorch_understanding_2024}, a tool that can disclose memory activities directly from the PyTorch CUDACachingAllocator. The results show that the error ranges from 0\% to a maximum of 1.2\%, supporting the use of gradient sizes to allocate all required memory blocks. Since the memory allocated for the model is persistent, deallocation times are unnecessary. Additionally, to resolve sequence mismatches, as backward propagation generates memory in reverse order of model loading, the allocation order is corrected accordingly.

\subsubsection{Batch Data} \label{design/chapter/sequence/batch-data}
refers to a subset of the dataset processed collectively in a single iteration. Consequently, memory blocks associated with batch data exist only for the duration of an iteration and are consistently relocated before the next one is loaded. The \projectname obtains batch data memory directly from the PyTorch dataloader \cite{pytorch_datasets_2024}. The time required for memory allocation and deallocation is actually the same as the duration of the iteration.

\subsubsection{Gradients} \label{design/chapter/sequence/grad}
refer to memory blocks allocated during backward propagation, utilized to minimize loss value during an optimization phase. These memory blocks are treated as temporary memory allocations when training on a CPU, implying that the memory blocks are released shortly after allocation. However, in GPU-based training, the release of this memory is controlled by the \texttt{optimizer.zero\_grad} function. To address this issue, the free time of all memory blocks for each layer is adjusted to the call time of the next \texttt{optimizer.zero\_grad} in iterations. In the absence of an \texttt{optimizer.zero\_grad} call in the following iteration, the memory blocks of all gradients remain allocated until \texttt{optimizer.zero\_grad} is eventually called.

\subsubsection{Optimizer} \label{design/chapter/sequence/optimiser}
refers to a collection of memory blocks produced during the execution span of an \texttt{optimizer.step}. Since optimizers such as Adam \cite{kingma_adam_2014} maintain extra memory blocks that have the same dimensions as the parameters of the model, \projectname excludes all memory blocks whose sizes do not match any of the model parameters' memory sizes. All memory blocks generated at this stage are labeled as permanent memory, lasting until the epoch is complete.

\subsubsection{Repetitive Iterations} \label{design/chapter/sequence/repetitive}
\projectname is capable of generating memory sequences for multiple iterations due to the repetitive iteration characteristics of DL \cite{222611, peng_optimus_2018, 227623, deep_gao_2022}. When the Stochastic Gradient Descent (SGD) \cite{ruder_overview_2016} optimizer is used, a memory sequence from a single iteration can accurately estimate the peak memory usage for the training task. However, for other optimizers like Adam, the first iteration often generates additional permanent memory blocks during the optimization phase, making it less accurate to predict peak memory usage based on the memory sequence of a single iteration. To accommodate a wide range of scenarios, xMen, by default, generates memory sequences for two iterations as input for the allocator.

\begin{figure}[tbp]
    \centering
    \includegraphics[width=\columnwidth]{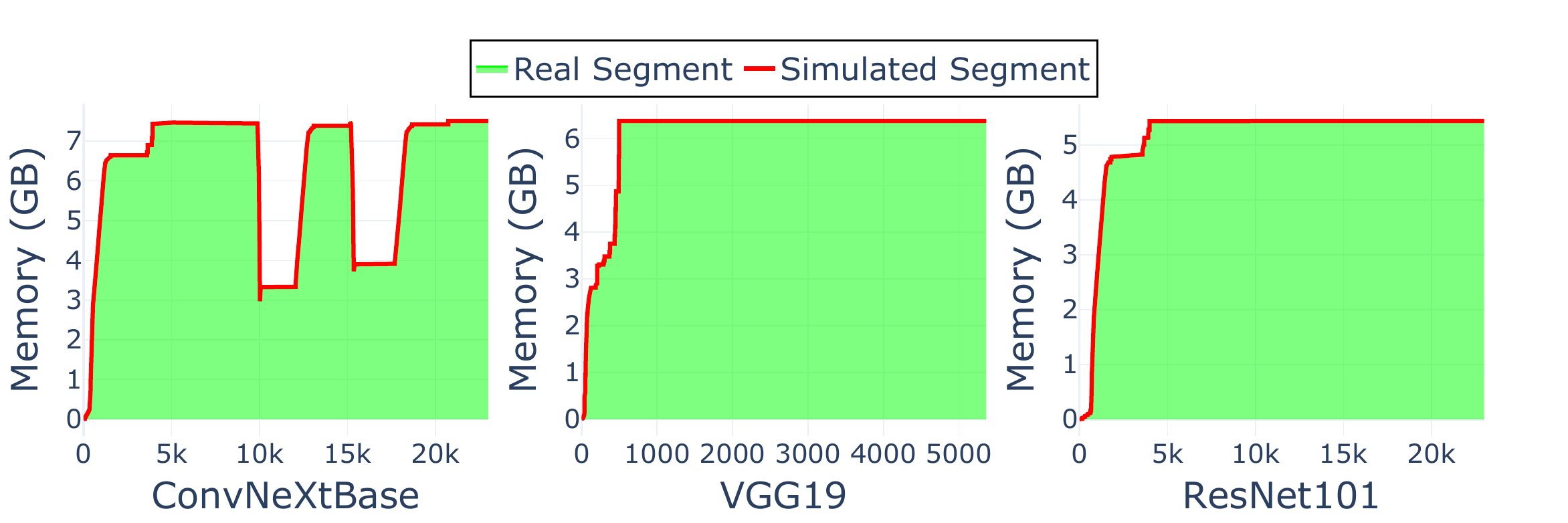}
    \caption{A comparison of memory allocation between the snapshot real and allocator simulation segments.}
    \label{design/fig/allocator-assessment}
\end{figure}

\ifthenelse{\boolean{isPreprint}}{%
}{
    \subsection{Memory Allocation Simulator} \label{design/chapter/memort-allocator}
    A memory allocator is a crucial in estimating the peak GPU memory usage. By default, PyTorch employs a CUDACachingAllocator for all GPU training tasks. Due to its inherent integration with the CUDA APIs, this allocator cannot be executed independently without a GPU. As a result, we reimplemented a caching allocator simulator in Python based on the C++ source code of the caching allocator \cite{pytorch_github-pytorch_2024}.
    
    The simulator employs four main logic components, as follows. 
    \begin{enumerate*}[label=(\roman*)]
        \item \textbf{Alignment}: rounds up the memory sizes of all requests to the nearest multiple of 512 bytes.
        \item \textbf{Segment size}: using Algorithm~\ref{background/alg/segment-size-calculation}, this logic determines the size of memory segments that are stored in the memory pool.
        \item \textbf{BFC}: a memory allocation algorithm used to search for an available memory block within a segment pool that best accommodates the present memory request. The memory pool is sorted using a lexicographic approach taking into consideration the CUDA stream, size of the memory block, and memory addresses. If requested memory size is smaller than the available memory block, the block is split, allocating the required portion to the request and returning the remainder to the pool. When a memory block is freed, the adjacent unallocated blocks are coalesced to form a larger block, which is then made available for future requests. 
        \item \textbf{Memory allocation mechanism}: a critical component of the simulator, dynamically manages the allocation and release of memory segments when both cache and GPU memory are exhausted. The BFC algorithm allocates a memory block from the existing pool of segments if a fitted memory block is found. Otherwise, it requests a new segment from the GPU and allocates it using the BFC algorithm. If no memory is available on either the GPU or the segment pool, it returns the segments to the GPU to free up memory first by releasing the segments over a threshold controlled by \texttt{max\_split\_size\_mb} \cite{pytorch_cuda_2024}, and then releasing all available segments.
    \end{enumerate*}
    
    To assess the precision of the segment activities generated by the simulator, we performed a comparative analysis against real-world data, namely the memory segment traces captured by a Snapshot Profiler for three distinct DL models. The results, illustrated in Fig.~\ref{design/fig/allocator-assessment}, exhibit an exact fit between the simulated (solid red line) and actual segment traces (green area). Note that the analysis focused only on the first 22,000 sampling points if total sampling points exceeded the number, as memory usage stabilized beyond this range.

}

%% file: evaluation.tex
\begin{table}[tbp]
\caption{The models and optimizers used in the experiments.}
\begin{center}
\begin{tabular}{cc|c}
\toprule
\textbf{Model} & \textbf{Release Year} & \textbf{Optimizers} \\
\midrule
VGG11 \cite{simonyan_very_2014} & 2014 &   \\
VGG16 \cite{simonyan_very_2014} & " &  \\
VGG19 \cite{simonyan_very_2014} & " &  \\
ResNet50 \cite{he_deep_2016} & 2016 &   \\
ResNet101 \cite{he_deep_2016} &  " &   \\
ResNet152 \cite{he_deep_2016} & " &  \\
MobileNetV2 \cite{sandler_mobilenetv2_2018} & 2018 & Adam \\
MobileNetV3(Small) \cite{howard_searching_2019} & 2019 & SGD \\
MobileNetV3(Large) \cite{howard_searching_2019} & " & Adagrad   \\
MnasNet \cite{tan_mnasnet_2019} & " & RMSprop  \\
RegNetX(400MF) \cite{radosavovic_designing_2020} & 2020 & AdamW  \\
RegNetX(32GF) \cite{radosavovic_designing_2020} & " &  \\
RegNetY(400MF) \cite{radosavovic_designing_2020} & " &  \\
RegNetY(32GF) \cite{radosavovic_designing_2020} & " &  \\
ConvNeXt(Tiny) \cite{liu_convnet_2022} & 2022 &  \\
ConvNeXt(Base) \cite{liu_convnet_2022} & " &  \\
\bottomrule
\end{tabular}
\label{evaluation/table/model}
\end{center}
\end{table}


This section evaluates the effectiveness of \projectname 
against three state-of-the-art baselines that employ distinct approaches to address the challenge of accurately estimating GPU memory usage. 
The evaluation was conducted through two separate experiments using the execution methodology described in Section~\ref{evaluation/chapter/methodology}.
The first experiment (\S\ref{evaluation/chapter/experiment-performance}) used the Analysis of Variance (ANOVA) method \cite{scheffe_analysis_1999} to analyze the relative error and examine the relationship between the probability of failed estimations, runtime, and median relative error. 
\ifthenelse{\boolean{isPreprint}}{%
}{
    The second experiment (\S\ref{evaluation/chapter/monte-carlo-experiment}) utilized the Monte Carlo method \cite{rubinstein_simulation_2017} to simulate a scenario in which a user submits tasks to a production cluster. This approach assesses the near-real performance for each estimator by conserving memory and calculating performance scores based on the median relative error and the probability of estimation failure.
}

\subsection{Baselines}
In establishing a robust foundation for comparative analysis, we carefully selected baselines from the literature encompassing diverse methodologies: static analysis, DL-based prediction, and direct GPU utilization. 
As a representative static analysis approach, we employed DNNmem \cite{gao_estimating_2020}, which combines a memory allocator with static analysis. 
While both SchedTune \cite{albahar_schedtune_2022} and TBEM \cite{liu_tbem_2022} use data-driven, pre-trained DL models for memory prediction, we selected SchedTune due to the availability of its source code, facilitating reproducibility. 
Finally, we included LLMem \cite{kim_llmem_2024} as a recent approach that utilizes direct GPU interaction. 
We deemed Horus \cite{yeung_horus_2022} less relevant to our comparative study owing to its primary focus being on FLOPS utilization and it only having a relatively simplistic memory estimation strategy. 
We refer the reader to Section~\ref{related-work/chapter} for a  detailed inspection of these and other related works.

For consistency, all evaluation plots use the following color scheme: {\color{exp-xMem}{blue}} for \projectname, {\color{exp-DNNMem}red} for DNNMem, {\color{exp-SchedTune}green} for SchedTune, and {\color{exp-LLMem}purple} for LLMem. Unless otherwise stated, `memory size' refers to the segment memory as described in Section~\ref{background/chapter/segment}.

\subsection{Experimental Setup}
The experiments were conducted on a server equipped with a 24-core Intel i9 CPU with 128GB RAM, and two GPUs: a GeForce 4070Ti with 12GB of memory and a GeForce 4060 with 8GB of memory. To ensure consistency and reliability for each experiment run, we utilized container technology with an official PyTorch image 
(\text{pytorch:2.3.1-cuda12.1-cudnn8-devel}) as the base to build an image of an experimental environment with all the necessary Python dependencies. 
During each run, a new container was initiated to ensure isolation and independence. In addition, each GPU was dedicated to a single run, ensuring that only one process fully occupied a GPU at any time. 
Additionally, we configured the CUBLAS workspace size and CUFFT plan size to zero within PyTorch to maintain consistent experimental conditions and avoid unintentional caching effects.

Table~\ref{evaluation/table/model} lists the models and optimizers used in our experiments, with a consistent input shape of [channels: 3, width: 86, height: 86]. The 4070Ti GPU was connected to a monitor that used 127MB of GPU memory, while a terminal for running the evaluation script required an extra 112MB, plus approximately 15MB of initial memory per GPU. Thus, each GPU had a fixed initial amount of memory in all the experiments, denoted as \(M^{\text{init}}_{d}\).


\ifthenelse{\boolean{isPreprint}}{%
    \begin{table}[tbp]
    \caption{Notations used in the evaluation experiments.}
    \begin{center}
    \rowcolors{2}{}{stripe}
    \begin{tabular}{>{\centering\arraybackslash}m{0.9cm} m{6.8cm}}
    \toprule
    \textbf{Notation} & \multicolumn{1}{c}{\textbf{Definition}} \\
    \midrule
    $j$ & One test configuration, including model, optimizer, \etc \\
    $i$ & Verification Round: 1\textsuperscript{st} or 2\textsuperscript{nd}\\
    $d$ & The index of the target GPU device, $d \in \{0,1\}$ \\
    $e$ & The estimator\\
    $N$ & The number of performed runs\\
    $M^{\text{init}}_{d}$ & The amount of memory used on device $ d$ for the duration of the experiment \\
    $M_{d}^{\text{max}}$ & The memory capacity of device $d$ \\
    $M^{\text{peak}}_{jid}$ & The peak memory usage as recorded by NVML during training with configuration $j$ on device $d$ at the $i$\textsuperscript{th} validation \\
    $\hat{M}^{\text{peak}}_{jde}$ & The peak memory usage as predicted by estimator $e$ using configuration $j$ and memory capacity of device $d$ \\
    $\hat{\text{OOM}}_{jde}$ & Boolean prediction of OOM occurrence by estimator $e$ using configuration $j$ on device $d$ \\
    $\text{OOM}_{jid}$ & Boolean indicating actual OOM occurrence during training with configuration $j$ on device $d$ at $i$\textsuperscript{th} validation \\
    $C_{jide}$ & Boolean indicating whether the prediction $\hat{\text{OOM}}_{jde}$ matches the actual $\text{OOM}_{jdi}$ at the $1$\textsuperscript{st} validation. In the $2$\textsuperscript{nd} validation, $C_{jde1}$ is utilized to further assess its estimation status. \\
    $\text{error}_{jide}$ & The error of $M^{\text{peak}}_{jid}$ relative to $\hat{M}^{\text{peak}}_{jde}$ \\
    $\tilde{\text{error}}_{jide}$ & The median of a set of $\text{error}_{jide}$  \\
    \bottomrule
    \end{tabular}
    \label{evaluation/table/notations}
    \end{center}
    \end{table}
}{
    \begin{table}[tbp]
    \caption{Notations used in the evaluation experiments.}
    \begin{center}
    \rowcolors{2}{}{stripe}
    \begin{tabular}{>{\centering\arraybackslash}m{0.9cm} m{6.8cm}}
    \toprule
    \textbf{Notation} & \multicolumn{1}{c}{\textbf{Definition}} \\
    \midrule
    $j$ & One test configuration, including model, optimizer, \etc \\
    $i$ & Verification Round: 1\textsuperscript{st} or 2\textsuperscript{nd}\\
    $d$ & The index of the target GPU device, $d \in \{0,1\}$ \\
    $e$ & The estimator\\
    $N$ & The number of performed runs\\
    $M^{\text{init}}_{d}$ & The amount of memory used on device $ d$ for the duration of the experiment \\
    $M_{d}^{\text{max}}$ & The memory capacity of device $d$ \\
    $M^{\text{peak}}_{jid}$ & The peak memory usage as recorded by NVML during training with configuration $j$ on device $d$ at the $i$\textsuperscript{th} validation \\
    $\hat{M}^{\text{peak}}_{jde}$ & The peak memory usage as predicted by estimator $e$ using configuration $j$ and memory capacity of device $d$ \\
    $M^{\text{save}}_{jde}$ & The memory conserved by estimator $e$ with configuration $j$ on device $d$ \\
    $\text{avgM}^{\text{save}}_{e}$ & The average memory conserved by running the estimator $e$\\
    $\hat{\text{OOM}}_{jde}$ & Boolean prediction of OOM occurrence by estimator $e$ using configuration $j$ on device $d$ \\
    $\text{OOM}_{jid}$ & Boolean indicating actual OOM occurrence during training with configuration $j$ on device $d$ at $i$\textsuperscript{th} validation \\
    $C_{jide}$ & Boolean indicating whether the prediction $\hat{\text{OOM}}_{jde}$ matches the actual $\text{OOM}_{jdi}$ at the $1$\textsuperscript{st} validation. In the $2$\textsuperscript{nd} validation, $C_{jde1}$ is utilized to further assess its estimation status. \\
    $\text{error}_{jide}$ & The error of $M^{\text{peak}}_{jid}$ relative to $\hat{M}^{\text{peak}}_{jde}$ \\
    $\tilde{\text{error}}_{jide}$ & The median of a set of $\text{error}_{jide}$  \\
    $P_{jie}$ & Probability of estimation failure by estimator $e$ with configuration $j$ at the $i$\textsuperscript{th} validation \\
    $PS_{ie}$ & The performance score of estimator $e$ at the $i$\textsuperscript{th} validation\\
    \bottomrule
    \end{tabular}
    \label{evaluation/table/notations}
    \end{center}
    \end{table}
}

\subsection{Methodology} \label{evaluation/chapter/methodology}
The notations used here are summarized in Table~\ref{evaluation/table/notations}.

Each run indicates a test configuration, $j$, performed once for the four estimators, denoted as $e$. The configuration comprises the model, optimizer, batch size, and when \texttt{zero\_grad} is called during training, at two positions: the first to call \texttt{zero\_grad} immediately before backward propagation and another at the start of each iteration. Furthermore, each run was divided into two validation steps, denoted as $i$, to evaluate the performance of each estimator separately:%
\begin{itemize*}[label=]
\item \textbf{Initial validation:} determines whether submitting tasks leads to an OOM error when training on a GPU with the maximum memory of the device; and%
\item \textbf{Subsequent validation:} focuses on the precision of the estimated peak memory and determines whether training causes OOM when using the estimated peak memory predicted by the estimator as the maximum runnable GPU memory.
\end{itemize*}

\subsubsection{Initial validation} 
aims to verify whether a predicted OOM result aligns with the actual OOM scenario and to compute the relative error, which follows these steps:
\begin{enumerate*}[label=(\roman*)]
    \item \textbf{Estimated peak memory}: predicts the peak GPU memory $\hat{M}^{\text{peak}}_{jde}$ during training based on a configuration $j$ and the maximum memory capacity of the device $d$.
    
    \item \textbf{Estimated OOM}: uses \eqref{evaluation/equation/estimation-oom} to determine the estimated $\hat{\text{OOM}}_{jde}$.
    
    \item \textbf{Actual OOM}: run task with configuration $j$ on device $d$ to determine\footnote{
    We limit each run to ten iterations, as peak memory tends to stabilize after a few iterations (\cf Section~\ref{design/chapter/sequence/repetitive}).\label{evaluation/foot/reason-speedup-evaluation}} whether training occurs $\text{OOM}_{jd1}$, expressed as \eqref{evaluation/equation/real-oom}. Meanwhile, NVIDIA NVML \cite{nvidia_corporation_nvml-nvidia_2024} measures the actual peak memory $M^{\text{peak}}_{jd1}$.
    
    \item \textbf{Estimation Correctness}: uses \eqref{evaluation/equation/1st-correctness-estimation} to verify the correctness $C_{jde1}$.
    
    \item \textbf{Relative error}: uses \eqref{evaluation/equation/relative-error} to calculate $\text{error}_{jde1}$.
\end{enumerate*}

\begin{equation}
    \label{evaluation/equation/estimation-oom}
    \hat{\text{OOM}}_{jde} = 
    \begin{cases} 
    1, &  \hat{M}^{\text{peak}}_{jde} > M_{d}^{\text{max}} \\
    0, &  \hat{M}^{\text{peak}}_{jde} \leq M_{d}^{\text{max}} \\
    \end{cases}
\end{equation}
\vspace{2mm}
\begin{equation}
    \label{evaluation/equation/real-oom}
    \text{OOM}_{jid} = 
    \begin{cases} 
    1, &  True \\
    0, &  False \\
    \end{cases}
\end{equation}
\vspace{2mm}
\begin{equation}
    \label{evaluation/equation/1st-correctness-estimation}
    C_{jde1} = 
    \begin{cases} 
    1, &  \hat{\text{OOM}}_{jde} = \text{OOM}_{jd1}, \\
    0, &  \hat{\text{OOM}}_{jde} \neq \text{OOM}_{jd1}.
    \end{cases}
\end{equation}
\vspace{2mm}
\begin{equation}
    \label{evaluation/equation/correctness-estimation/2nd}
    C_{jde2} = 
    \begin{cases} 
    1, &   C_{jde1}=1 \land \text{OOM}_{jd2} = 0 \\
    1, &   C_{jde1}=1 \land \text{OOM}_{jd1}=1 \\
    0, &  otherwise
    \end{cases}
\end{equation}
\vspace{2mm}
\begin{equation}
    \text{error}_{jide} = 
    \frac{\| \hat{M}^{\text{peak}}_{jde} - M^{\text{peak}}_{jid} \|}{M^{\text{peak}}_{jid}}
    \label{evaluation/equation/relative-error}
\end{equation}
\vspace{1mm}

\subsubsection{Subsequent validation} 
aims to determine whether training task with configuration \(j\) on target device \(d\) encounters an OOM issue when using the estimated peak memory predicted by the estimator as the maximum runnable GPU memory. Then, it verifies the accuracy of the estimated peak memory as follows:
\begin{enumerate*}[label=(\roman*)]
    \item \textbf{Set maximum memory}: uses $M^{\text{init}}_{d} + \hat{M}^{\text{peak}}_{jde}$ as maximum runnable memory to limit usage on the target device $d$. We use $M_{d}^{\text{max}}$ if this sum exceeds $M_{d}^{\text{max}}$.
    
    \item \textbf{Actual OOM}: run task with configuration $j$ on device $d$ to determine\footref{evaluation/foot/reason-speedup-evaluation} whether training occurs $\text{OOM}_{jd2}$, expressed as \eqref{evaluation/equation/real-oom} and measures actual peak memory $M^{\text{peak}}_{jd2}$ similarly to the initial validation.

    \item \textbf{Estimation Correctness}: uses \eqref{evaluation/equation/correctness-estimation/2nd} to verify the correctness $C_{jde2}$.

    \item \textbf{Relative error}: uses \eqref{evaluation/equation/relative-error} to calculate $\text{error}_{jde2}$.
\end{enumerate*}

\begin{figure*}[tbp]
\centering
\begin{subfigure}{2\columnwidth}
    \includegraphics[width=\linewidth]{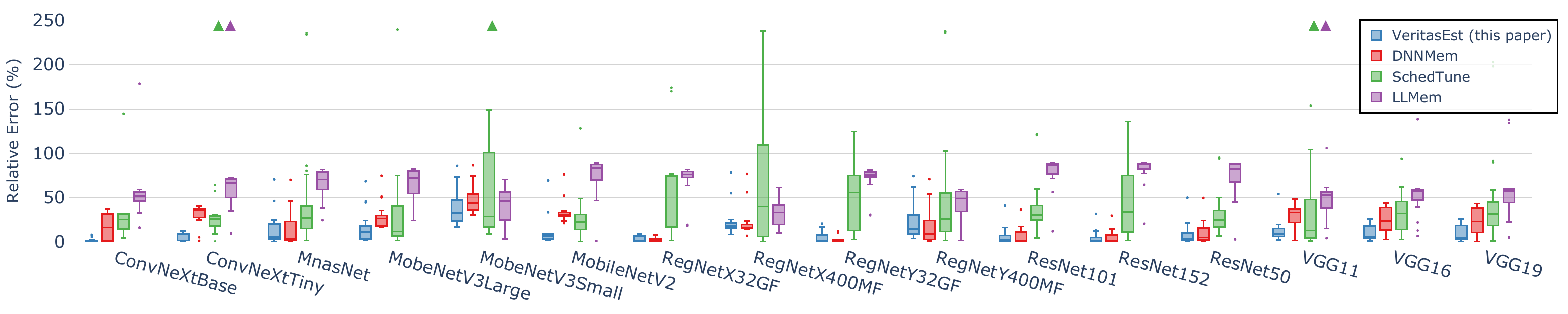}
    \caption{SGD optimizer}
    \label{evaluation/fig/relative-error-across-estimator/SGD}
\end{subfigure}
\vspace{1em}
\begin{subfigure}{2\columnwidth}
    \includegraphics[width=\linewidth]{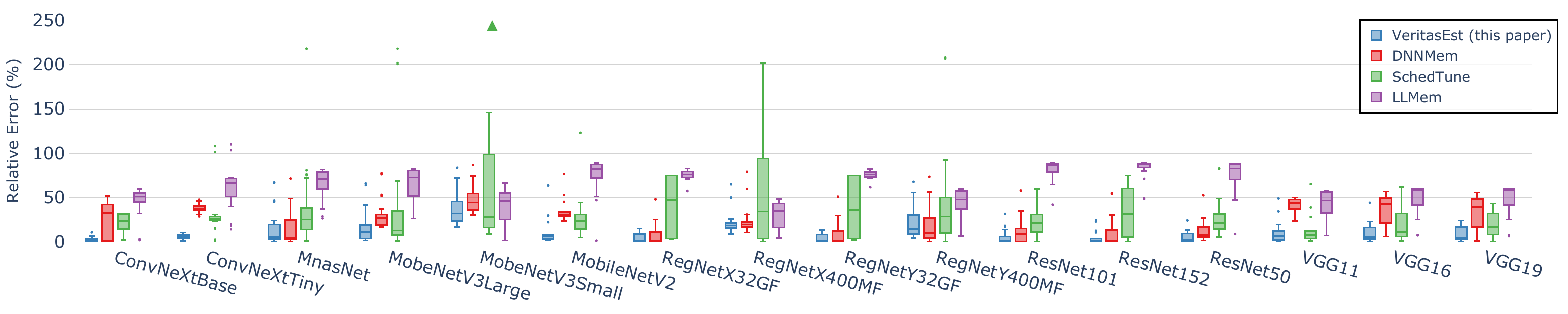}
    \caption{Adam optimizer}
    \label{evaluation/fig/relative-error-across-estimator/Adam}
\end{subfigure}

\caption{Relative error between predicted and actual memory usage in different models by estimators. Symbol $\blacktriangle$ denotes outliers exceeding 250\%; color follows the estimator legend.}
\label{evaluation/fig/relative-error-across-estimator}
\end{figure*}

\subsection{ANOVA Experiment} \label{evaluation/chapter/experiment-performance}
We paired all models and optimizers (Table~\ref{evaluation/table/model}) with 16 different batch sizes [10, 530, step: 40] to form 1,120 unique experimental configurations. Each configuration was repeated three times on an NVIDIA GeForce 4060 to minimize the impact of outliers (only 38 failures from a total of 3,360 runs). Each outcome contained four evaluation results generated by the three baselines and \projectname{}. Data were categorized into five groups, with each group representing results derived from training with a particular optimizer. Due to space limitations, we focus on two optimizers, SGD and Adam, which were also used as default optimizers in the evaluation sections of other baselines \cite{gao_estimating_2020, albahar_schedtune_2022, kim_llmem_2024}. The results from the other three optimizers were very similar to those obtained with Adam.

\subsubsection{Relative error} \label{evaluation/chapter/mean-relative-error}
The plots in Fig.~\ref{evaluation/fig/relative-error-across-estimator} illustrate \(\text{error}_{jde1}\) across various models, typically comprising 42 records per box. 
Fig.~\ref{evaluation/fig/relative-error-across-estimator/SGD} shows the relative error when SGD is used as an optimizer in a training task. \projectname{} consistently outperforms the baselines, having a lower median relative error in 8 of 16 models. The error for \projectname{} ranges from 0.17\% to 32.56\%, with an overall median error of 6.68\%. The most competitive baseline is DNNMem, 
yet \projectname{} achieves a lower overall median error compared to that of DNNMem (16.76\%). In contrast, SchedTune and LLmem exhibit notable error variability in this experiment, with SchedTune's maximum error reaching 387.44\% and LLmem's 306.69\%. 

Fig.~\ref{evaluation/fig/relative-error-across-estimator/Adam} is the equivalent with SGD as the default optimizer. Again, \projectname maintains its performance with an overall median error of 5.93\%. This is attributed to the evaluation of its dynamic memory mechanism throughout the training, which enables the capacity to capture the memory that Adam introduces during the optimization process accurately. In contrast, DNNMem has worse accuracy at 23.60\% owing to it employing static memory analysis. Furthermore, SchedTune maintains its high variability in this experiment, with a maximum relative error of 278.48\%. However, LLmem's maximum relative error decreases to 110.03\%.

\subsubsection{Probability of failed estimation} \label{evaluation/chapter/probability-preventing-oom}

\begin{figure}[t]
\centering
\begin{subfigure}{0.492\columnwidth}
    \includegraphics[width=\textwidth]{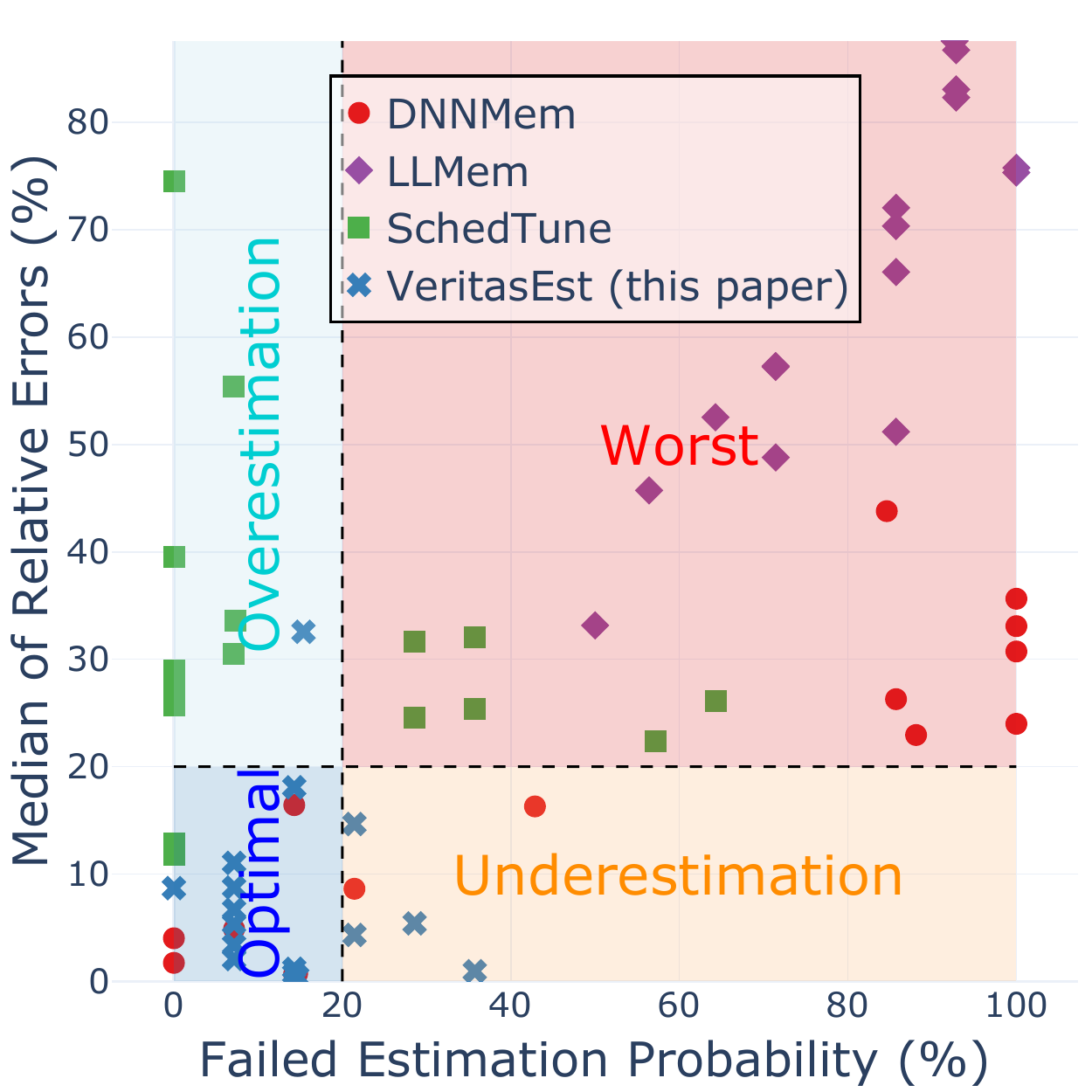}
    \caption{SGD optimizer}
    \label{evaluation/fig/probability-vs-error/SGD}
\end{subfigure}
\hfill%
\begin{subfigure}{0.492\columnwidth}
    \includegraphics[width=\linewidth]{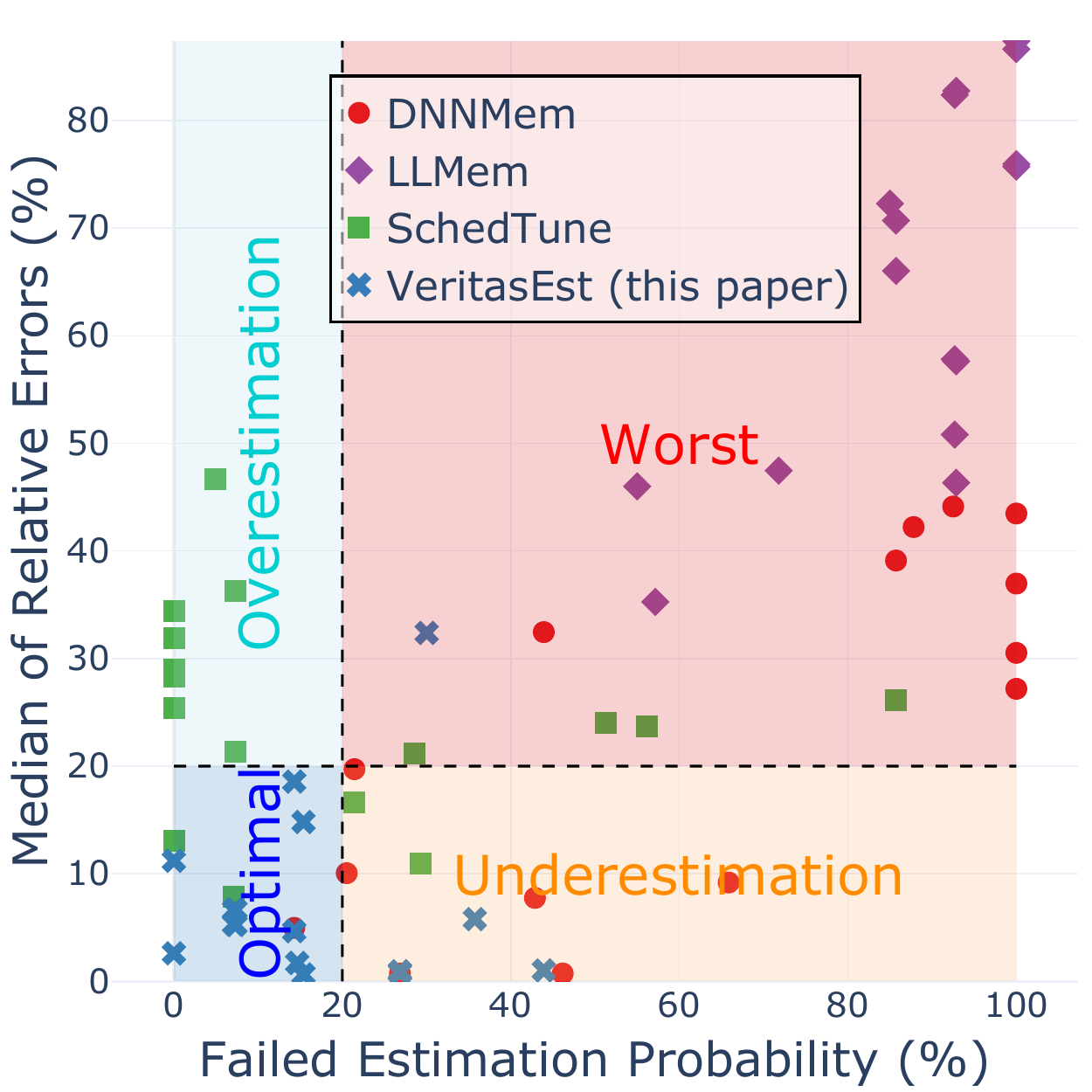}
      \caption{Adam optimizer}
      \label{evaluation/fig/probability-vs-error/Adam}
\end{subfigure}
\caption{Four-quadrant analysis of estimator performance through failed estimation probability and median relative error. Quadrants: Bottom left is {\color{fq-bl}optimal}; bottom right {\color{fq-br}underestimation}; top left {\color{fq-tl}overestimation}; and top right {\color{fq-tr}worst}.}
\label{evaluation/fig/probability-vs-error}
\end{figure}

Since the relative error does not indicate the probability of estimation failure for the estimator, this section uses Fig.~\ref{evaluation/fig/probability-vs-error} to depict the relationship between the probability of estimation failure ($P_{je2}$) on the x-axis, as defined in \eqref{evaluation/equation/failed-estimation-probability}, and the median relative error ($\tilde{\text{error}}_{jde2}$) on the y-axis, as defined in \eqref{evaluation/equation/median-error}. Each marker represents 42 results.

\begin{equation}
    \label{evaluation/equation/failed-estimation-probability}
     P_{jie}= \frac{N-\sum_{n=1}^{N} C_{jide}}{N}
\end{equation}
\vspace{1mm}
\begin{equation}
    \label{evaluation/equation/median-error}
    \tilde{\text{error}}_{jide} = \mathrm{median}(\text{error}_{jide})
\end{equation}
\vspace{1mm}


We chose 20\% as the threshold value for both axes to divide the plot into quadrants in order to distinguish acceptable estimation performance. A median relative error and failure probability below 20\% are considered satisfactory, whereas those above indicate predictive risk. Each quadrant represents a distinct classification or condition:
\begin{enumerate*}[label=(\roman*)]
    \item \textbf{Bottom left} is the \textit{optimal} quadrant, indicating that the results in this region have a reduced probability of estimation failure, with the estimated peak memory close to the actual value. 
    \item \textbf{Bottom right} is the \textit{underestimation} quadrant, where the results show a high probability of failure due to insufficient estimated peak memory for model training. In other words, the estimator predicts peak memory well but is not suitable to avoid OOM. 
    \item \textbf{Top left} represents the \textit{overestimation} quadrant, where results in this area have a lower probability of estimation failure, despite exhibiting a high relative error. Although it can potentially prevent OOM to some extent, results in this quadrant often inaccurately assess a model in an OOM condition, even when memory is available during training.     
    \item Finally, \textbf{Top right} features the \textit{worst} results, \ie those with both high relative error and high probability of estimation failure.
\end{enumerate*}

\ifthenelse{\not\boolean{isPreprint}}{%
    \begin{figure*}[t]
    \centering
    \begin{subfigure}{.5\columnwidth}
      \centering
      \includegraphics[width=\linewidth]{Evaluation-Comparison of Success Rate vs Relative Error-Monte Carlo.pdf}
      \caption{Four-Quadrant analysis}
      \label{evaluation/fig/probability-vs-error/monte-carlo}
    \end{subfigure}
    \begin{subfigure}{0.5\columnwidth}
      \centering
      \includegraphics[width=\linewidth]{Evaluation-CDF of Performance Score for Monte Carlo Experiment.pdf}
      \caption{Performance score CDF}
      \label{evaluation/fig/performance-score-cdf}
    \end{subfigure}
    \begin{subfigure}{1\columnwidth}
      \centering
      \includegraphics[width=\linewidth]{Evaluation-Accumulated Memory Saved by Estimator in Monte Carlo Experiment.pdf}
      \caption{Average memory conservation}
      \label{evaluation/fig/conserved-memory-by-estimator}
    \end{subfigure}
    \caption{Performance of estimators in Monte Carlo runs in three plots: failure probability and median error (Fig.~\ref{evaluation/fig/probability-vs-error/monte-carlo}); performance scores in initial and subsequent validations (Fig.~\ref{evaluation/fig/performance-score-cdf}); and average memory conserved per execution (Fig.~\ref{evaluation/fig/conserved-memory-by-estimator}).}
    \label{evaluation/fig/monte-carlo-results}
    \end{figure*}
}

Fig.~\ref{evaluation/fig/probability-vs-error/SGD} presents the results of the four estimators using the SGD optimizer. 
Overall, \projectname performs robustly in terms of median relative error and estimation failure probability. 
Most of the results of \projectname are in the optimal quadrant, with four in the underestimation quadrant and one in the overestimation. 
By contrast, SchedTune's results predominantly lie in the overestimation quadrant, which can be attributed to its notable error variability. DNNMem, while exhibiting excellent performance in the preceding performance experiment, shows polarized results, with part of its results falling in the optimal quadrant, and seven results positioned in the worst quadrant including four that demonstrate a prediction failure rate of 100\%. Finally, LLMem, consistent with its performance in the relative error experiment, shows unsatisfactory results.

Fig.~\ref{evaluation/fig/probability-vs-error/Adam} depicts the results obtained using the Adam optimizer, which are not dissimilar to those in Fig. ~\ref{evaluation/fig/probability-vs-error/SGD}. Most of the \projectname results remain in the optimal quadrant, except for one point that lies in the worst quadrant rather than the overestimation quadrant. Despite this, it is still closer to the origin than many other points. 
SchedTune and LLMem exhibited results very similar to those with SGD. DNNMem, on the other hand,  demonstrated substantial differences; the results that were originally in the optimal quadrant underwent a pronounced rightward shift, leaving only one point in the optimal quadrant. The majority of the remaining data points migrated to the underestimation and worst quadrants, owing to the limited insights made available through static analysis.

\subsubsection{Runtime}
The average runtime of \projectname (10.98 seconds) is longer than LLMem (5.40 seconds) and SchedTune (1.55 seconds), but DNNMem is the slowest (61.08 seconds). 
This performance difference stems from the fact that both \projectname and DNNMem employ data analytical methods to predict memory usage, leading to an increased processing time as the number of layers increases. In contrast, SchedTune achieves the fastest runtime by employing pre-trained models for direct estimation, and LLMem's runtime is only slightly slower than that of SchedTune because it gathers memory usage data directly using GPUs. 
\ifthenelse{\not\boolean{isPreprint}}{%
    Due to space limitations, the runtime plots are omitted but can be accessed in their entirety on GitHub\footref{introduction/foot/source-code}. 
}

\ifthenelse{\not\boolean{isPreprint}}{%
    \subsection{Monte Carlo Experiments} \label{evaluation/chapter/monte-carlo-experiment}
    We employed the Monte Carlo method \cite{kroese_why_2014} to thoroughly evaluate \projectname's 
    efficacy particularly in accurately estimating memory to avoid OOM and reduce memory usage. We also applied this method to the three baselines. 
    To follow the methodology, each execution randomly constructed a training configuration using the models and optimizers listed in Table~\ref{evaluation/table/model}, with batch sizes ranging from [1, 1000], GPUs selected from [Geforce 4060, GeForce 4070 Ti], and \texttt{zero\_grad}, as described in Section~\ref{evaluation/chapter/methodology}. Subsequently, this execution was repeated 1,680 times to assess the overall performance of each estimator and conserved memory by successfully predicting OOM.
    
    Using \eqref{evaluation/equation/performance-score-estimator} to calculate the performance score \(\text{PS}_{ie} \), two factors were considered: $P_{jie}$ and $\tilde{\text{error}}_{jide}$. 
    For predicting GPU memory usage to prevent OOM, prioritizing accuracy to prevent OOM is more important than minimizing prediction error. Thus, $P_{jie}$ is weighted (\(W_p\)) at 0.7, while a 0.3 weight (\(W_e\)) is assigned to $\tilde{\text{error}}_{jide}$, indicating the performance of predictions that were already successful. 
    Since both $P_{jie}$ and $\tilde{\text{error}}_{jide}$ range between 0 and 1, further normalization is unnecessary. 
    \begin{equation}
        \label{evaluation/equation/performance-score-estimator}
        \text{PS}_{ie} = W_p \times P_{jie} + W_e \times \tilde{\text{error}}_{jide}
    \end{equation}
    \vspace{1mm}
    
    \subsubsection{Performance}
    The \projectname results\footnote{Runtime box and scatter plots are omitted due to space, but are available on the repository\footref{introduction/foot/source-code}.} 
    demonstrate that the median relative error is 4.20\%, the probability of estimation failure is 8.21\%, the mean runtime is 13.19 seconds, the performance score with the initial validation results is 0.11, and with the subsequent validation results is 0.17. In this section, we utilize quadrant and CDF plots to assess the performance of the four estimators in the Monte Carlo experiment.
    
    Fig.~\ref{evaluation/fig/probability-vs-error/monte-carlo} shows that the results for \projectname, SchedTune, and DNNMem fall within the optimal quadrant. However, \projectname has only two results outside the optimal quadrant: one located in the overestimation quadrant and the other in the underestimation quadrant, both of which are near the optimal quadrant. In contrast, DNNMem had 11 results outside the optimal quadrant including five in the worst quadrant, whereas SchedTune had only four outcomes in the optimal quadrant. LLMem's performance consistently aligns with earlier findings in that all its results land in the worst quadrant.
    
    In terms of performance scores, Fig.~\ref{evaluation/fig/performance-score-cdf} presents
    \begin{enumerate*}[label=(\roman*)]
        \item the \textbf{first validation} results (dashed lines), namely \(\tilde{\text{error}}_{jde1}\) and \(P_{je1}\). \projectname achieves the best performance, with 95\% of its scores below or equal to 0.11. DNNMem ranked second, with 95\% of its scores below or equal to 0.15, followed by SchedTune, with 95\% of its scores below or equal to 0.23. In contrast, LLMem performs poorly, with 95\% of its scores at 0.82, which is significantly higher than those of the other three estimators.
        \item The \textbf{second validation} results (solid lines), namely \(\tilde{\text{error}}_{jde2}\) and \(P_{je2}\), indicate that the CDF curve of \projectname shifts rightward by 0.06, reaching a 95\textsuperscript{th} percentile of 0.17. SchedTune exhibited a substantial shift of 0.16, exceeding that of \projectname by a factor of more than two. DNNMem demonstrated the largest shift (0.48), which was six times greater than the \projectname shift. Meanwhile, LLMem consistently exhibited poorer performance with minimal enhancement.
    \end{enumerate*}
    
    \subsubsection{Memory savings} \label{evaluation/chapter/monte-carlo/memory-saving}
    We used equation \eqref{evaluation/equation/memory-savings} to evaluate memory savings in each estimation. Notably, we imposed a penalty indicating the negative amount as memory savings, when an estimator provided an incorrect estimation, implying that the estimator wasted a training task. Subsequently, equation \eqref{evaluation/equation/estimator-memory-average-saving} was used to calculate the average memory savings achieved per estimator when the peak memory was correctly estimated. Since there is a penalty, the memory savings might be negative, indicating the amount of wasted memory. 
    Fig.~\ref{evaluation/fig/conserved-memory-by-estimator} depicts the amount of memory conserved (y-axis) across models (x-axis). xMem demonstrated superior performance in 12 of the 16 evaluated models, with SchedTune marginally outperforming it in three instances. 
    Only in the case of MnasNet does xMem's performance fall slightly short of that of DNNmem. Average memory conservation reveals that \projectname consistently achieves mean memory savings of 5.69 GB per execution. SchedTune follows as the second-best performer with 4.87 GB, followed by DNNMem with 4.66 GB. Conversely, LLMem yields a value of -2.13 GB, underscoring significant memory inefficiency.
    
    \begin{equation}
        \label{evaluation/equation/memory-savings}
        M^{\text{save}}_{jde} = 
        \begin{cases} 
        M^{\text{max}}_d - \hat{M}^{\text{peak}}_{jde}, &   C_{jde1}=1 \land \text{OOM}_{jd2} = 0 \\
        M^{\text{max}}_d, &   C_{jde1}=1 \land \text{OOM}_{jd1}=1 \\
        -M^{\text{max}}_d, &  otherwise \\
        \end{cases}
    \end{equation}
    \vspace{2mm}
    \begin{equation}
        \label{evaluation/equation/estimator-memory-average-saving}
        \text{avgM}^{\text{save}}_{e} = \frac{\sum_{n=1}^{N} M^{\text{save}}_{jde}}{N}
    \end{equation}
    \vspace{1mm}

}

%% file: related_work.tex
Accurately measuring DL task requirements in the cloud is essential for preventing OOM errors and promoting efficient resource management. Numerous studies in the distributed systems literature have explored approaches such as static task requirement estimation, DL modeling, and resource scheduling to address the OOM issue. We now survey these.

\subsection{GPU Memory Estimator}
Recent studies have explored GPU memory estimation during model training from various perspectives. 
Xiao \etal \cite{222611} introduced a memory estimation method in Gandiva, employing online profiling to track GPU memory use and pinpoint peak and low usage moments. 
Gao \etal \cite{gao_estimating_2020} introduced DNNMem. It provides systematic estimates of GPU memory consumption by simulating memory-related activities in a memory allocator, derived from static analysis of the computational graph of DL models and user-defined cost functions. 
Horus~\cite{yeung_horus_2022} estimates GPU memory utilization considering four key factors: batch size, number of activations, gradients, and parameters, along with initialization overhead, to calculate the total memory requirement for a DL job. 
TBEM~\cite{liu_tbem_2022} estimates GPU memory consumption using a hybrid approach that combines the static analysis of computation graphs and dynamic testing in real environments with GPUs. 
In contrast, SchedTune~\cite{albahar_schedtune_2022} utilizes machine learning to estimate the maximum GPU memory requirements for DL tasks considering both model-specific features and GPU characteristics. 
Kim \etal \cite{kim_llmem_2024} proposed LLMem that estimates GPU memory usage by incorporating comprehensive evaluations of model architectures, fine-tuning strategies, and distributed computation dynamics, thereby facilitating precise peak memory predictions. 
Unlike \projectname, solutions such as DNNMem and TBEM overlook the effect of user code on memory sequences in their predictions, sticking to static computation graphs. This limits their ability to adjust to changes from user code modifications. Moreover, only DNNMem accounts for the impact of the memory allocator, as \projectname does. 

\subsection{OOM-awareness}
We now examine an alternative methodology for the prediction of GPU memory consumption to mitigate OOM issues.
The main work here is AntMan~\cite{258957}, which constructs a universal memory management layer based on GPU memory and RAM allocators. This allows AntMan to allocate partial model tensors into RAM if the GPU is running out of memory. This solution has been demonstrated to be highly effective for GPU sharing. However, if a training task itself exceeds the maximum memory capacity of the selected GPU, the portion of memory that exceeds the GPU limit will remain permanently in the RAM, leading to degraded training performance and wasted resources. By contrast, \projectname can accurately predict GPU memory requirements to determine whether the selected GPU is suitable for the current task in advance. Furthermore, \projectname can provide additional information on the resource management system for a more precise plan.

%% file: conclusion.tex
The toughest challenge in determining the memory consumption for model training is the scarcity of GPU resources. In this work, we reframed the initial research problem from \textit{``How to predict memory consumption 
with minimal GPU usage"} to \textit{``How to predict memory consumption without using a GPU at all."} This shift was driven by the realization that even minimal GPU usage for prediction purposes still leads to resource contention and inefficiencies. Prediction tasks submitted to GPU queues not only face delays due to limited availability but also contribute to congestion, increasing the load on the scheduler and complicating task allocation, thereby impacting the overall efficiency of resource management. This also leads to higher financial expenditures due to the cost of GPU resources.

Inspired by previous studies \cite{chen_slide_2019, daghaghi_accelerating_2021}, we propose \projectname as a tool that uses CPU-based analysis to predict the memory consumption required for model training on a GPU, allowing operation completely without a GPU. It addresses the aforementioned challenges, as CPUs are readily available and low-cost in both local and cluster environments. 
\projectname offers dual benefits for clusters: it eases cluster load by allowing users to precisely estimate memory consumption without resubmitting tasks, and it refines scheduling strategies by providing additional memory criteria to help increase resource allocation efficiency and prevent resource wastage on tasks prone to OOM errors. 

\ifthenelse{\boolean{isPreprint}}{
    The experiments indicate that our approach has a mere \xmemmedianerror median relative error with only a \xmemprobability chance of estimation failure and an average run time \xmemruntime. It also conserves an average of \xmemmemory of GPU memory from each estimation, significantly promoting GPU memory utilization. Notably, 95\% of its scores are 0.21 or lower when applying the second validation results. Compared to three state-of-the-art baselines, \projectname decreases the median relative error by \xmemmedianerrorimprove, the average probability of estimation failure by \xmemprobabilityimprove, and the runtime by \xmemruntimeimprove.
}{
    The experiments indicate that our approach has a mere \xmemmedianerror median relative error with only a \xmemprobability chance of estimation failure and an average run time \xmemruntime. It also conserves an average of \xmemmemory of GPU memory from each estimation, significantly promoting GPU memory utilization. Notably, 95\% of its scores are 0.21 or lower when applying the second validation results. Compared to three state-of-the-art baselines, \projectname decreases the median relative error by \xmemmedianerrorimprove, the average probability of estimation failure by \xmemprobabilityimprove, the runtime by \xmemruntimeimprove, and improves the capability of memory conserved by \xmemmemoryimprove. Importantly, \projectname achieved a 78.48\% reduction in the performance score with the initial validation result and a 69.58\% with the subsequent validation results, with a lower score indicating better performance.
}

Although PyTorch was used primarily in our experiments, the estimation methodology is applicable to other DL frameworks. Future work will address underestimations related to GPU cache data, improve \projectname's performance, and extend its capabilities to support distributed training and estimate memory usage for large language models (LLMs) during fine-tuning, all while maintaining GPU-free estimation.